\documentclass[aps,prd,preprintnumbers,showpacs]{revtex4}
\usepackage[dvips]{graphicx}
\begin{document}

%
%

\preprint{Nisho-10/1}
\title{ A Quintessence with Tiny Dark Energy Induced by Inflation }
\author{Aiichi Iwazaki} 
\affiliation{International Politics and Economics, Nishogakusha University, \\
Ohi Kashiwa Chiba 277-8585,\ Japan.} 
\date{Jun. 25, 2010}

\begin{abstract}
We propose a postulate $\dot{q} =MH$ concerning temporal variation of quintessence field $q$
where $H$ is Hubble parameter ( $=\dot{a}/a$ ) and $M$ is the scale characterizing the field.
Using the postulate we uniquely determines 
a dynamical model of the field. 
The potential of the field satisfies the tracking condition.
The minimum of  
the total energy density of matter and the field is located
at $a=\infty$. 
We show that
the present tiny dark energy 
is caused by early inflation, while the energy is
comparable to Planck scale before the inflation. 
Since  
the model is reduced to
$\Lambda$CDM model in the limit $M\to 0$,
it is a natural generalization of $\Lambda$CDM model.
\end{abstract}	
\hspace*{0.3cm}
\hspace*{1cm}

\maketitle
Recent observations\cite{1} have strongly suggested 
the acceleration of the cosmic expansion at present epoch.
The acceleration is caused by
the cosmological constant or a fluid of a scalar field with equation of state $\omega\simeq -1$.
It is an intriguing idea that the cosmological constant is
not constant but varying with time.
Hence, much attentions\cite{2} have recently been paid to such fluids describing 
the varying cosmological constant. There is, however, no guiding principle
for the construction of a promising model of the varying cosmological constant. 
Indeed, there is still not a promising model although many models with 
various potentials of the field have been proposed.
( Here we consider a real scalar field minimally coupled with gravity. )

Models of dynamical cosmological constant must satisfy observational constraints\cite{cons}, but
the constraints are not sufficiently stringent so as to uniquely determine a promising model.
Furthermore, there are serious problems to be solved. That is, we need to solve the problem why
the vacuum energy $\sim (10^{-3}\mbox{eV})^4 $ at present epoch is extremely smaller than 
the natural scale $\sim (\mbox{Planck mass})^4$. 
We also need to solve the problem why the cosmic expansion accelerates today.
At least, the expansion of the Universe does not accelerate 
around the epoch of
the recombination. It is a coincidence that we
observe the acceleration of the cosmic expansion at present epoch.
The solutions to these problems can't be found so far in previous models,
although the coincident problem is alleviated in models with tracking fields\cite{3}. 
( So called "tracking condition" for the potentials of the scalar field can be a guiding principle 
for the construction of 
models. But there are a variety of potentials satisfying the condition. Thus, it is
not restrictive so that a promising model is determined. )

In this paper
we propose a guiding principle for the construction of a specific model with
favorable properties.
The principle is in the following. A scalar field $q$ exists and it
satisfies a condition 

\begin{equation}
\label{1}
\frac{\dot{q}}{M}=\frac{\dot{a}}{a}=H,
\end{equation}
where $M$ denotes a scale characterizing the field and $H$ is Hubble parameter. 
( The dot denotes a derivative in $t$ and $a$ describes the scale factor of the flat Friedmann Universe;
$ ds^2=dt^2-a^2(t)d\vec{x}^2$. ) 
The principle should hold
in nature so that interactions of the scalar field with others
must preserve the condition. 
We do not know fundamental theoretical models with 
the field $q$ satisfying the condition. But postulating the existence of such a field,   
we obtain a promising model with no cosmological constant and coincidence problems.
Hence, we may regard the condition as a principle which must hold
in any fundamental models of the Universe.

Our model uniquely determined by the condition has remarkable properties.
The potential of the field is composed of two terms. One of them 
is an intrinsic for the field, while the other one
depends on the energy density of the background, i.e. radiations or non-relativistic
matter.
The minimum of total energy density of matter and the field is located at
$a= \infty$ and its value is equal to zero. 
The tiny dark energy at present epoch results from early inflation.
Namely, an exponential expansion of the Universe makes
the potential $V(q)$ of the field extremely small. Furthermore,
The equation of the state $w_q$ of the field is given by $w_q\simeq -1+4\lambda/3-z\lambda/(1+z)$,
where $z\lesssim O(1)$ denotes redshift and $\lambda$ is only adjustable small parameter in our model.
The model is reduced to $\Lambda$CDM model in the limit $\lambda \to 0$.

\vspace*{0.3cm}
Now, we show that our model of the scalar field $q$ is uniquely determined by the condition.
Hereafter, we assume that the field is spatially uniform but varying in time. 
First, 
we derive
the potential $V$ of the field.
Using the equation of motion $\ddot{q}+3H\dot{q}+\partial_qV=M\dot{H}+3MH^2+\partial_qV=0$,
we find that

\begin{equation}
\label{2}
\dot{q}(M\dot{H}+3MH^2+\partial_qV)=\frac{d}{dt}(\frac{M^2H^2}{2}+V)+3M^2H^3=0,
\end{equation}
where we used the condition $\dot{q}=MH$.

The Friedman equation is rewritten such that

\begin{equation}  
H^2=\frac{\rho_q+\rho_m}{m^2}=\frac{\dot{q}^2/2+V+\rho_m }{m^2}=\frac{M^2H^2/2+V+\rho_m}{m^2} ,
\end{equation}
with $m^2\equiv 3/8\pi G$,
where $\rho_q\equiv \dot{q}^2/2+V$ and $\rho_m$  
denote energy densities of the field and matter respectively. 
$G$ is gravitational constant. Thus, we obtain 

\begin{equation}
\label{H}
H^2=\frac{2(V+\rho_m)}{2m^2-M^2},
\end{equation}
where 
it is natural to take the scale $M$ be less than "Planck mass" $m\simeq 4.2\times 10^{18}$GeV.
Then, $H^2$ is positive. We notice that the total energy density $\rho_q+\rho_m$ can be expressed
in terms of the potential energy $V$ of $q$ and the energy density of matter $\rho_m$;
 $\rho_q+\rho_m \propto V+\rho_m$.
Inserting the formula in eq(\ref{H}) into the equation (\ref{2}), 
we obtain

\begin{equation}
\label{3}
2a\frac{d}{da}V+6\lambda V=-\lambda( a\frac{d}{da}\rho_m+6\rho_m ),
\end{equation}	
with $\lambda\equiv M^2/m^2<1$, where we used the relation $d/dt=\dot{a}\,d/da$. 

This is the equation to determine the potential $V$ when 
the energy density $\rho_m$ of matter is given as a function of $a$. 
For example in the matter dominated period 
when $\rho_m$ is given such as $\rho_m\equiv \tilde{\rho}_0a^{-3}$,
we find 

\begin{equation}
\label{ma}
V=\frac{\lambda \rho_m}{2(1-\lambda)}+\tilde{V}_0a^{-3\lambda}
\quad \mbox{in matter dominated period}
\end{equation}
where $\tilde{V_0}$ is a constant determined by the present values of $\Omega_q$ and $\Omega_m$
; $\Omega_q=\rho_q/(\rho_q+\rho_m)$ and $\Omega_m=\rho_m/(\rho_q+\rho_m)$.
Similarly, we obtain the potential in the radiation dominated period;
$V=\lambda \rho_m/(4-3\lambda)+\tilde{V}_0a^{-3\lambda}$ where $\rho_m\propto a^{-4}$.

\vspace*{0.2cm}
We notice that Hubble parameter $H$ or $V+\rho_m$ vanishes in the limit $a\to \infty$.
It implies that the total energy density of the matter and the field
vanishes in the limit. Since $H^2 \ge 0$, 
the minimum of the total energy density is located at $a=\infty$.
Even if we add a constant energy $E$ to $\rho_m$,
the solution $V(a)$ in eq(\ref{3}) has a term $-E$ so that the total energy density
proportional to $V+\rho_m$ does not change. Thus, the minimum
remains to be zero. This is a general result for 
energy density of matter with the behavior $\rho_m\propto a^{-n}$ ( $n>0$ ) as far as $\lambda\ll 1$.
( Indeed, the observations of $\omega_q(z\simeq 0)$ as well as 
a constraint derived from Big Bang Nucleosynthesis
require $\lambda\ll 1$, as we show below. )
Thus, our postulate in eq(\ref{1}) leads to the vanishing vacuum energy,
which is a prominent property in our model.

\vspace*{0.2cm}
The potential $V$ in eq(\ref{ma}) has been obtained as a function of the scale factor $a$ not the field $q$.
But it is easy to rewrite $V$ in terms of $q$.
Since we explicitly obtain $q=M\log (a/a_0)$ from the condition eq(\ref{1}), 
the potential $V$ can be rewritten such that

\begin{equation}
\label{p}
\bar{V}(q)\equiv V(a=a_0\exp(q/M))=\frac{\lambda \tilde{\rho}_0a_0^{-3}\exp(-\frac{3q}{M})}{2(1-\lambda)}
+\tilde{V}_0a_0^{-3\lambda}\exp(-\lambda \frac{3q}{M})
\quad \mbox{in matter dominated period},
\end{equation}
where $a_0$ is an integration constant and represents the period when
the field $q$ vanishes.

The potential $\tilde{V}_0a_0^{-3\lambda}$ intrinsic for $q$
should be of the order of $M^4$ when $q$ takes the value $M$. 
Thus, we may assume $\tilde{V}_0a_0^{-3\lambda}=M^4$. 
Then, it follows that

\begin{equation}
\label{V}
\bar{V}(q)=\frac{\lambda\rho_0(a/a_0)^{-3}}{2(1-\lambda)}+V_0(a/a_0)^{-3\lambda}
=\frac{\lambda \rho_0\exp(-\frac{3q}{M})}{2(1-\lambda)}
+V_0\exp(-\lambda \frac{3q}{M}) \quad \mbox{in matter dominated period},
\end{equation}
with $\rho_0\equiv \tilde{\rho}_0a_0^{-3}$ and $V_0\equiv \tilde{V}_0a_0^{-3\lambda}=M^4$.


\vspace*{0.3cm}
In this way we can find the explicit form of the potential $\bar{V}(q)$ by postulating
the condition. 
The potential is composed of two terms; 
one is the term depending on $\rho_m$ ( the first term in eq(\ref{V}) ) and 
the other one $V_{\rm{in}}\equiv \tilde{V}_0a^{-3\lambda}$ 
is the term intrinsic for the field $q$ ( the second term in eq(\ref{V}) ). 
The term depending on $\rho_m$ is smaller than $\rho_m$ itself since $\lambda \ll 1$
( see later ).
We point out that the intrinsic term $V_{\rm{in}}(a)$ of the potential decreases more slowly than
the first term depending on $\rho_m(a)$ or $\rho_m(a)$ 
as $a$ increases.  
Therefore, the total energy density $\rho_q(a)+\rho_m(a)$ eventually approaches $V_{\rm{in}}(a)$
with the expansion of the Universe.
We stress that with small $\lambda \ll 1$,
the potential in eq(\ref{V}) satisfies "tracking condition"\cite{trac} 
$\bar{V}''(q)\bar{V}(q)/\bar{V}'(q)^2>1$, where
$\bar{V}'\equiv \partial_q\bar{V}$ and $\bar{V}''\equiv \partial_q^2\bar{V}$.

The potential in eq(\ref{V}) has been discussed in previous papers \cite{dexp}, in which
more general ones than the potential $\bar{V}(q)$ have been considered 
such that $A\exp(-\alpha q/M)+B\exp(-\beta q/M)$. The parameters $A,B,\alpha$, $\beta$ and $M$ have been
partially determined so as to obtain quintessence with favorable "tracking fields".
But there are some freedoms of modifying these potentials\cite{mexp},
even if the consistency with observations is kept. On the other hand,
there are no such freedoms in our model;
if we add a potential $\delta \bar{V}$ to $\bar{V}$, 
the potential $\delta \bar{V}+\bar{V}$ does not satisfy the equation(\ref{3}) in general. 
Thus, the modification of the potential
contradicts with
the basic postulate eq(\ref{1}).
Furthermore,
we have only three parameters $M$, $\tilde{\rho}_0$ and $\tilde{V}_0$.
Among them,
$\tilde{\rho}_0$ and $\tilde{V}_0$ are determined with the present values of 
the density parameters $\Omega_q$ and $\Omega_m$. The remaining one $M$
is only adjustable parameter 
which can be determined by observing the equation of state of the field $q$ today.
Although our model involves only one adjustable parameter $M$ or $\lambda$, 
the model is shown to be consistent with current observations
when we choose small $\lambda \,\,(\,\,\ll 1$ ).

It is easy to see 
that the density parameters $\Omega_q$, $\Omega_m$ 
and the equation of state $\omega_q$ in the matter dominated period are respectively
given by,

\begin{eqnarray}
\label{omega-q}
\Omega_q&=&\frac{2V+\lambda\rho_m}{2(V+\rho_m)}=\frac{2(1-\lambda)\tilde{V}_0 a^{-3\lambda}
+(2-\lambda)\lambda\rho_m}{2(1-\lambda)\tilde{V}_0 a^{-3\lambda}+(2-\lambda)\rho_m}
\simeq \frac{2\tilde{V}_0 a^{-3\lambda}+2\lambda\tilde{\rho}_0a^{-3}}{2\tilde{V}_0 a^{-3\lambda}+2\tilde{\rho}_0a^{-3}},\\
\Omega_m&=&\frac{(2-\lambda)\rho_m}{2(V+\rho_m)}=\frac{(2-\lambda)(1-\lambda)\rho_m}{2(1-\lambda)\tilde{V}_0a^{-3\lambda}+(2-\lambda)\rho_m}
\simeq \frac{\tilde{\rho}_0 a^{-3}}{\tilde{V}_0 a^{-3\lambda}+\tilde{\rho}_0 a^{-3}},\\
\omega_q&=&\frac{p_q}{\rho_q}=\frac{\frac{\dot{q}^2}{2}-V}{\frac{\dot{q}^2}{2}+V}
=\frac{-2(1-\lambda)^2\tilde{V}_0 a^{-3\lambda}}{2(1-\lambda)\tilde{V}_0 a^{-3\lambda}+\lambda(2-\lambda)\rho_m}
\simeq \frac{-\tilde{V}_0 a^{-3\lambda}}{\tilde{V}_0 a^{-3\lambda}+\lambda\tilde{\rho}_0a^{-3}},
\end{eqnarray}
where we used $\dot{q}^2/2=\lambda (V+\rho_m)/(2-\lambda)$ and took the limit $\lambda \ll 1$.

Using the present values $\Omega_q=0.75$ and $\Omega_m=0.25$ at $a=1$ ( here we have taken
$a=1$ as the scale factor today ), we obtain the parameters
$\tilde{V}_0=0.75\tilde{\rho}_c$ and $\tilde{\rho}_0=0.25\tilde{\rho}_c$ where 
$\rho_c$ denotes the critical density $\tilde{\rho}_c\equiv \tilde{\rho}_0+\tilde{V}_0=\rho_0a_0^3+V_0a_0^{3\lambda}$ at present.
Furthermore, we find that 

\begin{equation}
\label{omega}
\omega_q(a\simeq 1)\simeq -1+\frac{4\lambda}{3}+\lambda (1-a), 
\end{equation} 
where we have taken the limit $a\to 1$.
Since the formula has only one parameter $\lambda$, 
we can check the validity\cite{w} of our model by observing
$\omega_q(a\simeq 1)$ in detail. 
As we have stated, current observations suggests 
$\omega_q(a=1)\lesssim -0.9$. It leads to the constraint $\lambda\lesssim 0.08$.

We can examine whether or not the field $q$ affects Big Bang Nucleosynthesis,
which occurs at about $a=10^{-9}$.
Since $\Omega_q\simeq \lambda$ for sufficiently small $a\ll 1$ 
such that $\lambda\, a^{-3}\gg a^{-3\lambda}$,
the field does not affect\cite{cons,n} the Nucleosynthesis
if $\lambda<0.1$. It is interesting that
both of these observations suggests the similar constraint $\lambda <0.1$.

We should mention that when we take the limit $M\to 0$ ( $\lambda\to 0$ ), 
the field $\dot{q}=MH$ vanishes and
the total energy density of the Universe becomes $\tilde{\rho}_0a^{-3}+\tilde{V}_0$ in the matter
dominated period; $\dot{q}^2/2+V(a)\to \tilde{V}_0$ in the limit $M\to 0$. 
Thus, our model
is reduced to the standard $\Lambda$CDM model with the cosmological constant $\Lambda=3\tilde{V}_0/m^2$.
( Here we take the limit 
keeping the critical density today $\tilde{\rho}_c=\tilde{\rho}_0+\tilde{V}_0$ fixed. )

It is instructive to see that the energy density of the field 
dominates the Universe only around the present
epoch $a\simeq 1$ since 

\begin{equation}
\frac{\Omega_q}{\Omega_m}=\frac{2(1-\lambda)\tilde{V}_0 a^{-3\lambda}
+(2-\lambda)\lambda\rho_m}{(2-\lambda)(1-\lambda)\rho_m}\simeq 
\frac{\tilde{V}_0 a^{-3\lambda}
+\lambda\rho_m}{\rho_m}=\frac{3 a^{-3\lambda}
+\lambda a^{-3}}{a^{-3}}
\end{equation}
with $\lambda\ll 1$, 
where $\rho_m=\tilde{\rho}_0a^{-3}$ 
and $\tilde{V}_0=3 \tilde{\rho}_0$.

We may estimate the turning point from the deceleration to the acceleration of the Universe, that is, 
the period when $\ddot{a}=0$.
The period is obtained by solving the equation $\ddot{a}\propto \rho_m+\rho_q+3p_q=0$.
Since it is rewritten as $2\tilde{V}_0/\tilde{\rho}_0\simeq (1+z)^3/(1+z)^{3\lambda}$ with $\lambda \ll 1$,
we find that 

\begin{equation}
\ddot{a}(z)=0 \quad \mbox{at} \quad z\simeq 0.94 \,\,\,(\, 0.83 \,)\quad \mbox{for} \quad \lambda=0.1 \,\,\,(\, \lambda=0.01 \,),
\end{equation}
where $z\equiv 1-1/a$ and $2\tilde{V}_0/\tilde{\rho}_0=6$. 
We note that $\ddot{a}=0$ at $z(a)\simeq 0.82$ in $\Lambda$CDM model, which can be obtained 
from our model by taking the limit $\lambda=M^2/m^2\to 0$.

\vspace*{0.3cm}
We have shown that using the condition in eq(\ref{1}) we uniquely determine 
a model of dynamical cosmological constant, which involves one free parameter $\lambda$
and is consistent with observations when $\lambda <0.1$. We have also shown that the model is reduced to
$\Lambda$CDM in the limit $\lambda \to 0$. 

We now proceed to show why the present value of dark energy 
$V(a=1)\simeq \tilde{V}_0=M^4a_0^{3\lambda}$ is much smaller than
the natural scale $m^4$ of the Universe. 
In other words, we show
why $a_0$ is extremely small such as
$a_0=(\tilde{V}_0/M^4)^{1/3\lambda}\sim 10^{-400}$ 
with $\tilde{V}_0=(10^{-3}\rm{eV})^4$ and $\lambda\sim 0.1$.
Obviously, the period $a=a_0$ when the field $q$ vanishes, 
is earlier than the period of the beginning of inflation\cite{inf}.
The inflation is needed for the scale factor to become so large that $a=1$ today.
We assume that classical treatment of the field is possible even in the period of inflation;
the field satisfies the condition in eq(\ref{1}) in the period of inflation.  
The inflation makes the scale factor $a$
exponentially increase with time. Thus, 
while it takes the value $M^4$ at $a=a_0$
before the inflation,
the inflation 
makes the dark energy  $V_{\rm{in}}\propto a^{-3\lambda}$
exponentially small, 
so that the present small dark energy is realized; 
$V_{\rm{in}}(a)=\tilde{V}_0a^{-3\lambda}=M^4(a_0/a)^{3\lambda}\to \tilde{V}_0$ as $a\to 1$.

In order to show how it can be actually realized in our model, 
we may explicitly take $\rho_{\rm{inf}}=ka^{\epsilon}$ as a flat potential for inflaton\cite{inf}
with $0<\epsilon \ll 1$. Then, $\rho_{\rm{inf}}$ is almost constant in $a$ ( $k$ is a constant
of the order of $M'^4$ where $M'$ is a typical scale of inflaton. )
Here, the precise form of the inflaton potential is not necessary for the later discussion.
Only what we need is the flatness of the potential in $a$. 
So, the parameter $\epsilon$ must be extremely small.

Assuming that the total energy density at this epoch is given by  
the potentials of the field $q$ and inflaton,
we find the potential $V(a)$ by
solving eq(\ref{3}) with $\rho_m=\rho_{\rm{inf}}$,

\begin{equation}
V=-\frac{k(1+\frac{\epsilon}{6})}{1+\frac{\epsilon}{3\lambda}}a^{\epsilon}+V_0(a/a_0)^{-3\lambda}
\simeq -\frac{k}{1+\frac{\epsilon}{3\lambda}}a^{\epsilon}+V_0(a/a_0)^{-3\lambda}=
 -\frac{k}{1+\frac{\epsilon}{3\lambda}}a^{\epsilon}+V_{\rm{in}}(a)
\end{equation}
where we use that $\epsilon \ll 1$. Thus, Hubble parameter is given by

\begin{equation}
H^2=\frac{2(V+\rho_{\rm{inf}})}{2m^2-M^2}
\simeq (1+\frac{\lambda}{2})m^{-2}\Bigl(\frac{k\epsilon}{3\lambda (1+\frac{\epsilon}{\lambda})}a^{\epsilon}+V_0(a/a_0)^{-3\lambda}\Bigr).
\end{equation}
where the first term in the right hand side of the equation 
represents the contribution of inflaton and the second term
does the contribution of the field $q$.
The second term decreases much faster than the first one
with the expansion of the Universe. Hence,
the first term eventually dominates the Hubble parameter $H$. Then, the inflation begins
since $H$ is almost constant in $a$ since $\epsilon \ll 1$.
Once it starts, the term $V_{\rm{in}}(a)=M^4(a_0/a)^{3\lambda}$
rapidly decreases with time.
Therefore, we have extremely smaller intrinsic potential 
$V_{\rm{in}}(a=a_f)=M^4(a_0/a_f)^{3\lambda}$ at 
the end $a=a_f (\gg a_i$) of the inflation than the potential
$V_{\rm{in}}(a=a_i)=M^4(a_0/a_i)^{3\lambda}$ at the beginning $a=a_i$ of the inflation.
In other words the dark energy can be much small at the end of the inflation,
although it is not necessarily small at the beginning of the inflation.
In this way 
we obtain much smaller dark energy at present than the natural scale $m^4$ of the Universe.


We may estimate e-folding number of the inflation leading to the small dark energy today.
Since the present value $\tilde{V}_0$ is of the order of $(10^{-3}\rm{eV})^4$,
the e-folding number $N$ of the inflation is roughly given by

\begin{equation}
\label{ef}
N=\log\frac{a_f}{a_i}\sim 920, \quad \mbox{since} \quad 
\frac{V_{\rm{in}}(a_i)}{V_{\rm{in}}(a_f=1)}
=\Bigl(\frac{a_i}{a_f=1}\Bigr)^{-3\lambda}=\Bigl(\frac{a_i}{a_f=1}\Bigr)^{-0.3}
\sim \Bigl(\frac{M}{10^{-3}\rm{eV}}\Bigr)^4\sim 10^{120},
\end{equation} 
where we have assumed $\lambda=M^2/m^2=0.1$. Obviously, this e-folding number is larger
than the actual one because the actual inflation ends 
at the period much earlier than today $a_f=1$; the period $a_f=1$ 
was assumed as the period of the end of the inflation in the above estimation.
The above estimation is an example to show that the inflation really makes the potential much small
by taking an appropriate value of the e-folding number N. 
In this way we can have much small intrinsic potential energy at the end of the inflation.
( The reason why $a_0$ is much small compared with the present value $a=1$
is obvious. Such a small $a$ is inflated to a large one and subsequently reaches $a=1$ 
with the standard expansion of the Universe. ) 

Finally, we obtain an upper bound of $N$ by considering the coincidence problem.
The bound comes from the fact that inflation with too large $N$ makes the potential $V_{\rm{in}}(a)$ become too small. Namely, the inflation with too large e-folding number N makes the 
density parameter $\Omega_q(a=1)$ of the dark energy at present
much smaller than the one of the dark matter.  
First,
we remind you of the sequence of events. Namely, the inflation begins at $a=a_i$ 
after the period $a=a_0$ when $q$ vanishes. Then, 
after the end $a=a_f$ of the inflation, rethermalization occurs at $a=a_{th}$.
Thus, we have the inequality that
$a_0<a_i<a_f=a_ie^N<a_{th}$ where the period $a_{th}$ denotes
the epoch when the Universe is rethermalized with temperature $T_r$.
For example, $T_r=10^{12}$ GeV is realized at $a_{th}=10^{-25}$.
Thus, it follows from the sequence that $N<\log(a_{th}/a_0)$, that is, 

\begin{equation}
N<(37\log 10+1.3\log\bar{M})/\lambda-29\log 10-\log(T_{th}/10^{16}\mbox{GeV}).
\end{equation}
It is upper bound for the e-folding number $N$.
For instance, $N<848-\log (T_{th}/10^{16}\mbox{GeV})$ in the above example with
$\lambda=0.1$, where $\bar{M}\simeq 1.3\times 10^2$.

\vspace*{0.3cm}

We obtain the equality today of the dark energy $V(a=1)\simeq \tilde{V}_0$
and dark matter $\rho_m(a=1)$
by taking 
a very small value $a_0=(\tilde{V}_0/M^4)^{1/3\lambda}$, e.g. $\sim 10^{-400}$.  
It apparently seems a fine tuning that 
we adjust the scale factor $a_0$ to such a small value. But,
the appearance of such a small value is the result obtained from
the formula $a_0=(\tilde{V}_0/M^4)^{1/3\lambda}$. What we do indeed is that we
adjust the parameters $M$ or $\lambda$ and $\tilde{V}_0$ to fit the observation of $\omega_q$
and $\Omega_q(a=1)$, respectively.
Obviously, such an adjustment is not fine tuning. Furthermore,
we have argued that even if such a small scale factor appears, 
the inflation with an appropriately chosen e-folding number N
makes this small scale factor $a$ become large.
Consequently we have $a=1$ today. Thus,
it is not so unnatural that we have such a small scale factor $a_0$.
Therefore, the cosmological constant and coincidence problems are solved in our model.

\vspace*{0.3cm}

To summarize, assuming the presence of a real scalar field
satisfying the condition $\dot{q}/M=H$,
we have derived a simple analytical model of a dynamical cosmological constant.
The model has several remarkable features.
The total energy density of the field and matter in the model is controlled by the field $q$ so that
the minimum of the energy is equal to zero. It is located at $q=+\infty$ or $a= \infty$. 
The energy density of the field is comparable to the natural scale $m^4$ in the epoch before
inflation.  But,
the inflation makes the energy density of the field extremely small.
As a result it becomes comparable to the density of the dark matter today.
Hence, the cosmological constant problem can be naturally solved.
Since there is only one adjustable parameter $\lambda$ involved in the model,
we can check a consistency of our model 
by observing the equation of state $\omega_q(a\sim 1)$ at present.
Furthermore, the model is reduced to the $\Lambda$CDM model in the limit $\lambda\to 0$.
Thus, our model is a natural generalization of the $\Lambda$CDM model.
Since our model is very satisfactory, 
we need to find theoretical basis for the postulate
with which we derive the model.

\end{document}